\documentclass[]{aa}    
\usepackage{natbib}
\bibpunct{(}{)}{;}{a}{}{,}

\input{psfig.sty}

\begin{document}

\title{INTEGRAL observations of the Crab pulsar}
\author{T.Mineo\inst{1}, C.Ferrigno\inst{1},  L.Foschini\inst{2},
 A.Segreto\inst{1}, G.Cusumano\inst{1}, G.Malaguti\inst{2},
 G.Di Cocco\inst{2}, C.Labanti\inst{2} }
\institute{INAF IASF-Pa, via U. La Malfa 153, 90146 Palermo,
  Italy \and
INAF IASF-Bo, via P.Gobetti 101, 40129 Bologna, Italy 
}

\offprints{T. Mineo: teresa.mineo@ifc.inaf.it}
\date{Received: ..... ; accepted: ......}
\titlerunning{INTEGRAL observation of the Crab pulsar}
\authorrunning{T. Mineo et al.}

\abstract{}{
The paper presents the timing and spectral analysis of several
observations  of the \object{Crab} pulsar performed  with INTEGRAL in the energy range
3-500 keV. }
{All these observations, when summed together provide a high
statistics data set which can be used for accurate phase resolved
spectroscopy. 
A detailed study of the pulsed emission at different phase
intervals is performed. }
{The  spectral distribution changes with phase showing a
characteristic reverse S shape of the photon index.
Moreover the spectrum  softens with energy, in each phase interval, 
and this behavior is adequately modeled over 
the whole energy range 3-500 keV
with a single curved law with a slope variable with Log(E), 
confirming the BeppoSAX results on the curvature of the pulsed emission.
The bending parameter of the log-parabolic model is compatible with a single
value of 0.14$\pm$0.02 over all phase intervals. }
{Results are discussed
within the three-dimensional outer gap model.}

\maketitle


\section{Introduction}
The Crab pulsar (\object{PSR B0531+21}) can be observed in
almost every energy band of the electromagnetic spectrum. Its pulse profile is
characterized by a double peak structure with a phase separation of 0.4 that is
almost aligned in absolute phase over all  wavelengths 
\citep{rots04,kuip03,tenn01}.
 
\begin{figure*}[t]
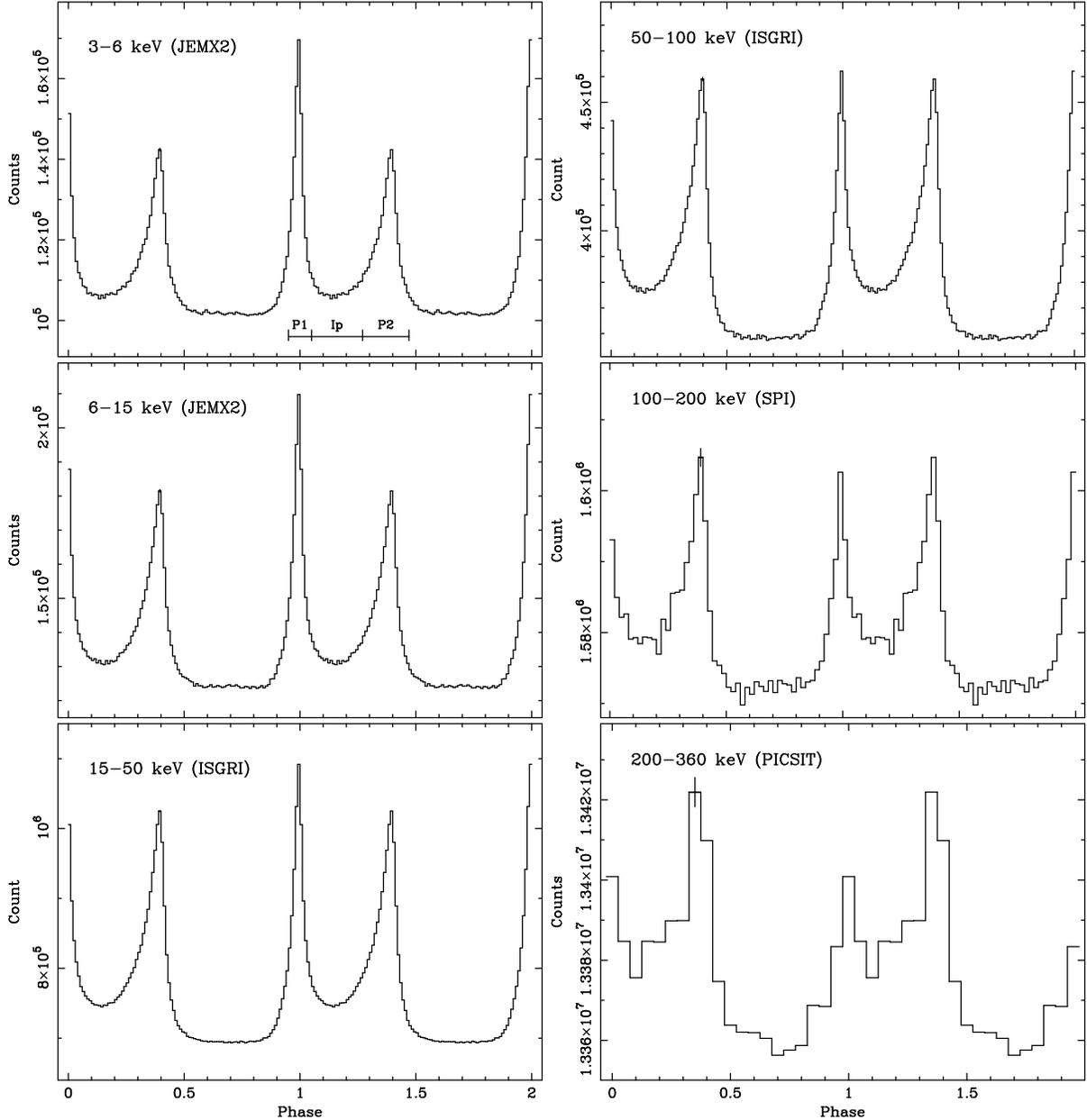

\label{fig1}
\centerline{ \vbox{
\hbox{
\psfig{figure=4305fi1a.ps,width=7.9cm,angle=-90,clip=} 
\psfig{figure=4305fi1b.ps,width=7.9cm,angle=-90,clip=}}
\hbox{
\psfig{figure=4305fi1c.ps,width=7.9cm,angle=-90,clip=} 
\psfig{figure=4305fi1d.ps,width=7.9cm,angle=-90,clip=}}
\hbox{
\psfig{figure=4305fi1e.ps,width=7.9cm,angle=-90,clip=} 
\psfig{figure=4305fi1f.ps,width=7.9cm,angle=-90,clip=}}
}}
\caption{Crab phase histograms in six energy bands in absolute phase 
(the main radio pulse at phase 0.0).
The light curves have different phase resolution according to the statistics 
available and to the time resolution of the instrument. 
P1, P2 and Ip phase intervals according to the definition 
given in \citet{mine97} are indicated in the top left panel.}
\end{figure*}

In the X-ray range, the relative intensity, height and width of the two
peaks vary with energy: 
the first peak (P1), dominant at low X-ray energies, becomes smaller
than the second one (P2). 
Moreover, an enhancement with energy of the
bridge between these peaks, usually called Interpeak (Ip), is also 
well evident \citep[][ see also Fig.1] {mine97}. 
At energy  above 1 MeV, the morphology changes abruptly:
the first pulse becomes again dominant over the second one 
and the bridge emission loses significance; the pulse profile 
above 30 MeV is similar to the one observed at optical wavelengths
\citep{kuip01}.
\\
A first detailed study of the phase-resolved X-ray spectra  
has been performed by \citet{prav97}, in the 5-200 keV
energy interval, based on RXTE (PCA and HEXTE) data. Their main result was
a variation of the photon index as function of the pulse phase with a
reverse S shape: the spectrum softens starting from the leading edge of the
first peak where it reaches the maximum value, it hardens in the interpeak
and softens again in the second peak.
The S shape spectral variation with phase has been reported  by \citet{mass00}
from  BeppoSAX data and by \citet {weis04} from Chandra data, 
even if with lower statistical significance, 
confirming the symmetric evolution of the spectral 
index around the first peak. However,
the softening of the P1 core respect to the leading edge has been recently
questioned by \citet{vive02} in a new analysis of  RXTE data  but performed over the
smaller energy range of 5-60 keV. 
\\
Significant X-ray emission from the pulsar in the 
off-pulse interval (phase 0.5-0.9) was discovered by \citet{tenn01}
with Chandra observations, however, the 
spectral index measured in this phase interval suffers of large statistical
uncertainty  \citep{weis04}.

BeppoSAX observations of the Crab pulsar showed that 
the photon indices of the pulsed emission significantly
increase with energy maintaining  the same S shape behavior 
over the 0.1-300 keV energy range \citep{mass03, zhan02}. 
The spectral index variation has been modeled using  a single
curved power law with a slope variable with Log(E) \citep{mass00}.
Moreover, applying this model to three
wide phase intervals, the first peak, the Interpeak  and the second peak, 
a single value of $\sim$ 0.15 for the curvature parameter has been 
measured in the three intervals \citep{mass01}. 
\\
\citet{kuip01} presented a coherent high-energy picture of the Crab
pulsar from 0.1 keV up to 10 GeV by using the high energy $\gamma$-ray data
from the CGRO satellite together with data obtained at  X-ray
energies from several observatories.
The authors model the 0.1 keV-10 GeV pulsed emission 
in 7 narrow phase slices with a composite model: a power law 
present in the phase intervals of the two main pulses, 
a curved spectral component required in the same 
phase intervals
and second broader curved spectral component representing mainly 
the bridge emission. 

X-ray observations of Crab  pulsar  performed 
with a balloon born experiment report the detection of an emission
line at 440 keV  with a flux of 
(0.86$\pm$0.33)$\times$10$^{-4}$ ph cm$^{-2}$ s$^{-1}$ \citep{mass91}.
\citet{ulme94}, using OSSE data, did not detected this line 
but derived a  3 $\sigma$ upper limit compatible with its presence.

\begin{table*}
\label{tab1} 
\caption{Observation log for the data used in this analysis}
\newcommand{\m}{\hphantom{$-$}}
\newcommand{\cc}[1]{\multicolumn{1}{c}{#1}}
\renewcommand{\tabcolsep}{0.7pc} 
\renewcommand{\arraystretch}{1.2} 
\begin{center}
\begin{tabular}{@{}ccrrrr}
\hline \hline
Revolution & \multicolumn{1}{c}{Start-Stop time} & \multicolumn{4}{c}{Exposure (ks)}\\ 
                  & (MJD)                  & JEM-X & ISGRI & PICsIT &SPI \\
\hline
39   &  52677.2 - 52679.8  & 149.3  & 142.3 & -- &--\\
40   &  52680.2 - 52681.6  & 122.6  & 56.0  & 106.5  &--\\
41   &  52683.2 - 52685.8  &  --  & -- & 204.0   &-- \\
42   &  52686.4 - 52688.2  & 75.4   & 61.6  & -- &138.5\\
43   &  52689.6 - 52691.7  & --   & 30.7  & -- & 154.9\\
44   &  52692.2 - 52694.4  & --  & 9.8  &-- &146.9\\
45   &  52695.2 - 52696.7  & --  & 2.1 & -- &27.4\\
102  &  52866.3 - 52868.1  & --  & 37.9 &  -- &78.0\\
103  &  52868.6 - 52868.8  &     & 21.6 & -- &--\\
\hline
\multicolumn{2}{l}{Total}  & 347.3   & 362.0 & 310.5  & 545.7\\                  
\hline
\end{tabular}
\end{center}
\end{table*}
\normalsize

Results on the INTEGRAL observations of the Crab pulsar 
have already been presented by \citet{kuip03},
that studied the instrument absolute timing accuracy
and by \citet{bran03} 
who reports results on the 3-37 keV energy range with the X-ray monitor JEM-X.

In this paper, we present the timing and spectral analysis of 
several observations of the Crab pulsar performed with 
SPI, JEM-X, IBIS/ISGRI and IBIS/PICsIT on board INTEGRAL. 
We investigate the presence of the 440 keV line and thanks to the wide
energy range covered by INTEGRAL instruments and to the good statistics
achieved by the summed data sets, we are able to 
perform a detailed  phase resolved spectroscopy on the Crab pulsed
emission over the wide energy range 3-500 keV.  

\begin{figure}[ht]
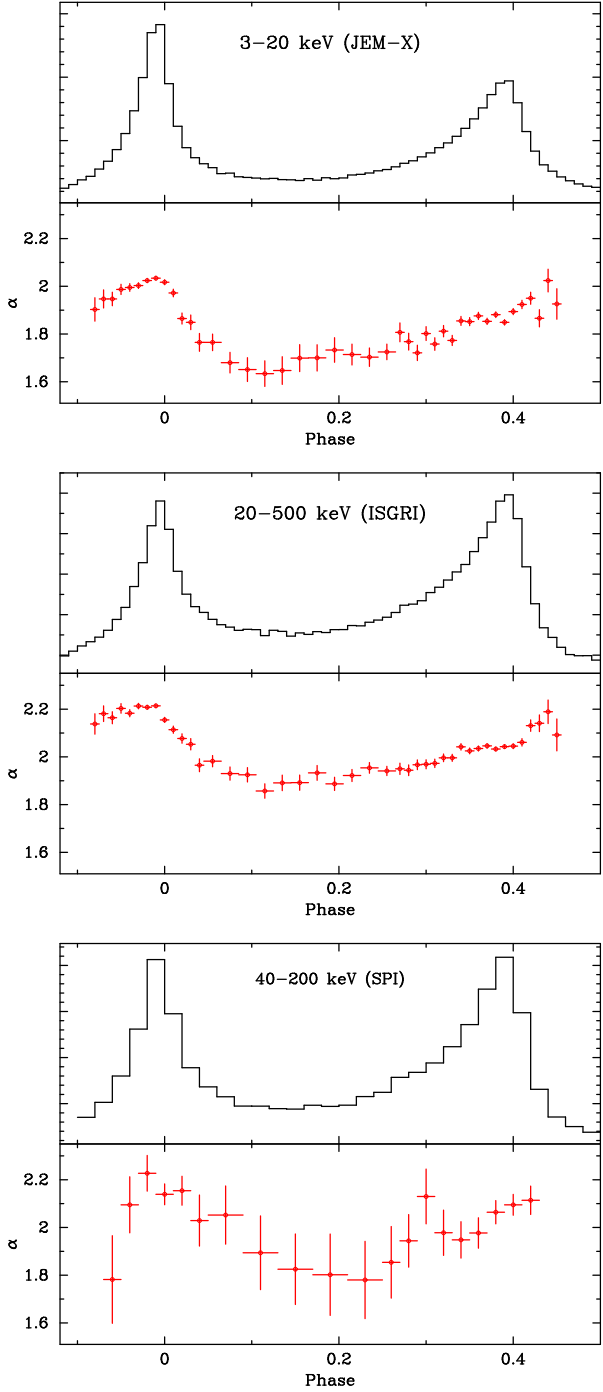

\label{fig2}
\centerline{ 
\vbox{
\psfig{figure=4305fi2a.ps,width=7.9cm,angle=-90,clip=}
\vspace{0.3 cm}
\psfig{figure=4305fi2b.ps,width=7.9cm,angle=-90,clip=}
\vspace{0.3 cm}
\psfig{figure=4305fi2c.ps,width=7.9cm,angle=-90,clip=}
}}
\caption{Spectral index vs. phase measured by JEM-X in the energy range 3-20
keV (top panel),  IBIS/ISGRI in the energy range 20-500 keV (middle panel)
and SPI in the energy range 40-200 keV (bottom panel).
}
\end{figure}


\section{Observation and Data reduction}
The International Gamma-Ray Astrophysics Laboratory \citep[INTEGRAL;][]{wink03}
observed the Crab nebula and pulsar for calibration purposes several
times from February 2003 (rev. 39) to August 2003 (rev. 103).

The INTEGRAL payload is composed of three  high energy instruments. 
JEM-X \citep{lund03} consists of two identical coded-aperture mask telescopes with a
geometrical area of 500 cm$^2$ and an angular resolution of 3 arcmin across an
effective field of view of about 10\degr.
The detector at the focal plane, a Microstrip Gas Chamber,  operates in the energy
range 3-35 keV with an energy resolution of $\sim$17\% at 6 keV. 
SPI \citep{vedr03} is a high spectral resolution gamma-ray telescope 
that consists of an array of 19 closely packed germanium detectors surrounded
 by an active anticoincidence shield of BGO. The imaging capabilities of the 
instrument are obtained with a tungsten coded aperture mask
adopting a particular observing strategy (dithering).
The fully coded field-of-view is 16\degr, and the angular resolution is 2.5\degr. 
The energy range extends from 20 keV to 8 MeV with a typical energy 
resolution of 2.5 keV at 1.3 MeV.
IBIS \citep{uber03} is a  coded aperture telescope composed by two detection layers:
ISGRI \citep{lebr03} and PICsIT \citep{dico03}. 
ISGRI is a large CdTe gamma-ray camera  operating in the range 15 keV--1 MeV,
with a geometrical area of 2621 cm$^2$ and an energy resolution of $\sim$8\% at
60 keV.   PICsIT is composed by 64$\times$64 Caesium-Iodide (CsI)
scintillation pixels working in the energy intervals 175 keV -- 10 MeV.

The limited telemetry budged of INTEGRAL  degrades  the event
timing {\bf information}  on board. Taking into account all possible 
uncertainties affecting the time accuracy (the On Board Time (OBT)
  accuracy, the orbit
prediction etc.), \citet{kuip03} estimated that 
the resulting time resolution is $90\mu$s for IBIS, $130\mu$s for SPI and $150\mu$s
for JEM-X,  about 30-40\% worse  than the nominal time resolution
of each single instrument; moreover
the INTEGRAL absolute timing accuracy, as estimated by \citet{kuip03}
from Crab data, is about 40$\mu$s.
The IBIS/PICsIT detector cannot be routinely configured in
photon-by-photon mode due to
the high telemetry budget requested for this operational mode.
 For timing studies then, observers can select the
spectral-timing mode in
which the whole detector counts are accumulated on board in up to eight
energy bands (default 4) and with an integration time in the range
0.97-500 ms (default 3.9 ms).

\begin{figure}[ht]
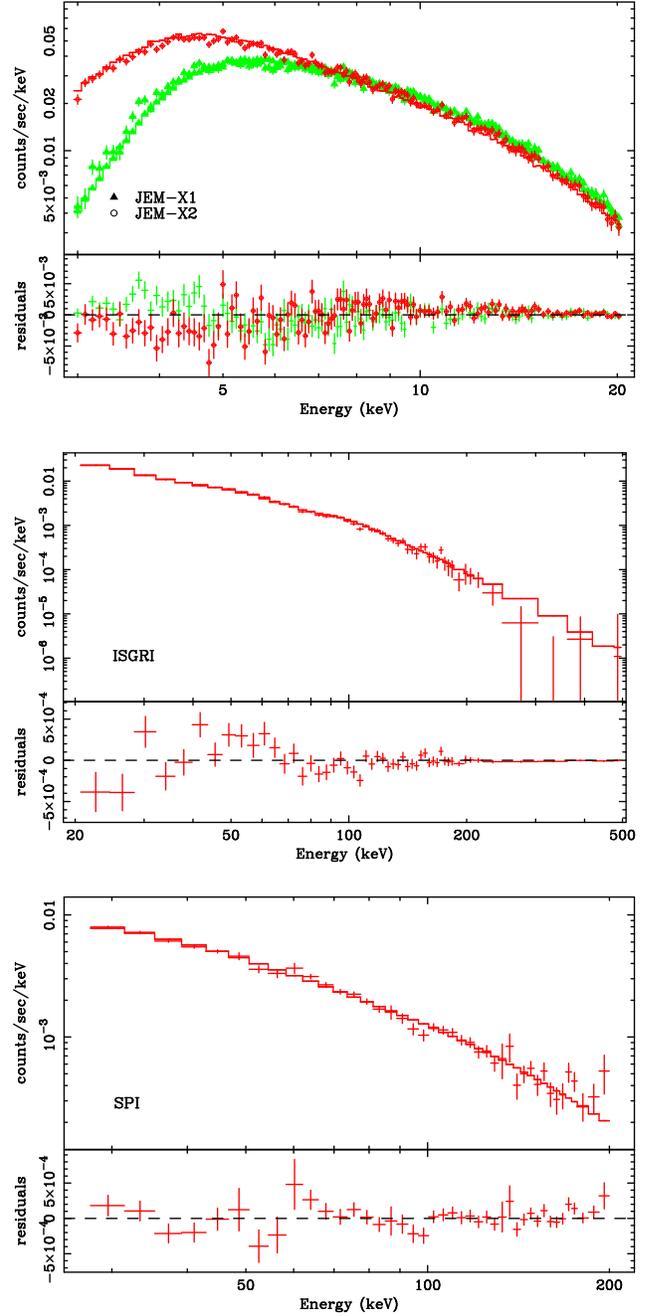

\label{fig3}
\centerline{ 
\vbox{
\psfig{figure=4305fi3a.ps,width=8.3cm,angle=-90,clip=}
\vspace{0.4 cm}
\psfig{figure=4305fi3b.ps,width=8.3cm,angle=-90,clip=}
\vspace{0.4 cm}
\psfig{figure=4305fi3c.ps,width=8.3cm,angle=-90,clip=}
}}
\caption{ JEM-X,  IBIS/ISGRI and SPI spectrum of the phase interval 0.99-1.00 fitted
  with a single power law}
\end{figure}
  
The analysis performed in this paper uses on axis JEM-X observations
and  SPI  data relative to pointings less than 6\degr.  
IBIS observations have a maximum off-axis angle of 1\degr. They are relative
to  different configuration of the instrument:
  in particular, data are accumulated on board with different
 rise-time selections. We then verified that spectra relative to different
science windows have negligible differences in  the energy
range considered for the spectral analysis. 
IBIS/PICsIT observation intervals with
time resolution of 1 ms. have been considered and 
 to  improve  the statistics of the
 light curve,  data from  rev. 0041 with an off-axis of 9.6\degr  
~were also included.
\\
Table 1 summarizes the log of the observations used in
this analysis together with the relative time exposures. 

To obtain spectra and light
curves of the  sources present in the field,  INTEGRAL official 
software (OSA\footnote{Available at \\
\texttt{http://isdc.unige.ch/index.cgi?Soft+download}}),
whose algorithms are described in \citet{gold03} and \citet{gros03} for IBIS, 
\citet{skin03},  and \citet{stro03} for SPI, and \citet{west03} for JEM-X,
repeats a shadowgram
deconvolution process several times, by selecting events in the energy and
time intervals of interest.
However, when the source positions have already been determined 
(because a priori known  or predetermined by a shadowgram deconvolution), 
it is  alternatively possible, to select only the detector pixels fully illuminated by the
 source. This method simplifies the accumulation of phase resolved spectra.
The amount of
illumination from a given source, normalized to the maximum illumination value is
called  Photon Illumination Fraction (PIF) and
 is generated by the standard software for JEM-X and IBIS/ISGRI. 
For these instruments, we run the standard pipeline (OSA vers. 4.2)
up to the ``DEAD'' level that includes the conversion from detector energy
 channels (PHA) to energy channels corrected for instrumental effects (PI),
 the selection for the Good Time Intervals (GTI) and the correction
for the instrument dead time and we selected  events with
 PIF$ = 1$. 
However, IBIS/ISGRI conversion PHA-to-PI has been performed 
through our own calibration file  generated by one of the authors (A.S.).
This file is based on on-ground and in-flight
calibration data and  represents an
improvement respect to the standard one (see Appendix A).

No PIF selection is possible for SPI;  we then extracted the 
list files relative to  the whole field of view  running the standard pipeline 
(OSA vers. 4.2) up to the ``COR'' level that produces corrected 
events selected for the GTI.

The response matrices used in the analysis of JEM-X and SPI data are provided 
by the standard software.  The ISGRI  response matrix  has been 
generated with our own software to take into account the new PHA-to-PI calibration file.
(see Appendix A).

 No response matrix is available 
 for spectral timing data with IBIS/PICsIT; the presently available
matrices are in fact suitable only for spectral imaging data. 
Data from this
detector have not been included in the spectral analysis. 

JEM-X spectral analysis was performed in
the energy range 3-20 keV, 
IBIS/ISGRI and SPI spectra were fitted in the range  20-500 keV and
40-200 keV, respectively. 
\\
Errors quoted in the paper are relative to 1 $\sigma$
confidence level for one interesting parameter. 

\begin{figure}[htb]
\label{fig4}
\centerline{ 
\hbox{
\psfig{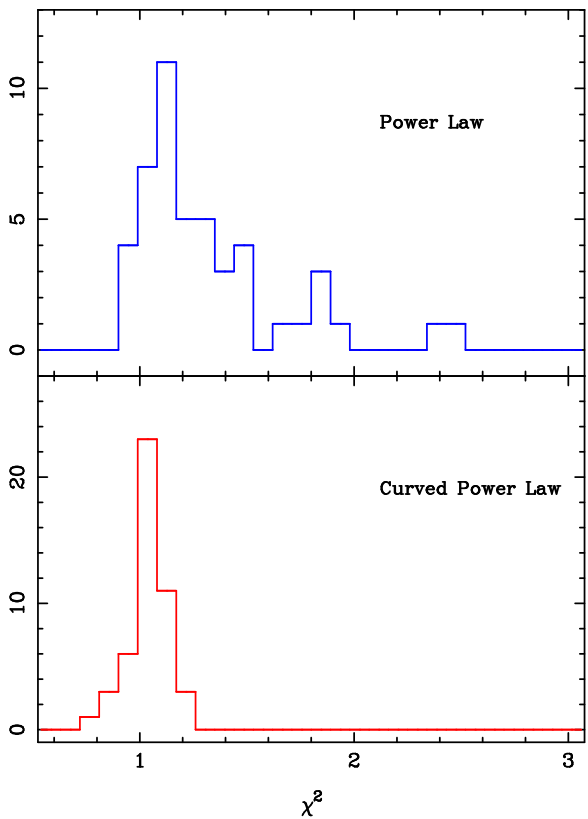} 
}}
\caption{Frequency histogram of the reduced $\chi^2$  for 256 dof obtained fitting 
simultaneously JEM-X2 and
IBIS/ISGRI spectra with a single power law (top panel) and with 
the curved model of Eq. (1)}
\end{figure}


\section{Timing Analysis}
Arrival times were converted to the Solar System Barycentre with the
 DE200 ephemeris. The values of $P$ and $\dot{P}$ in GRO format 
for each observation  (our data set spans several months)
were derived from Jodrell Bank Crab Pulsar Monthly Ephemeris
(http://www.jb.man.ac.uk/) using contemporary radio ephemeris.

\begin{figure}[htb]
\label{fig5}
\centerline{ 
\hbox{
\psfig{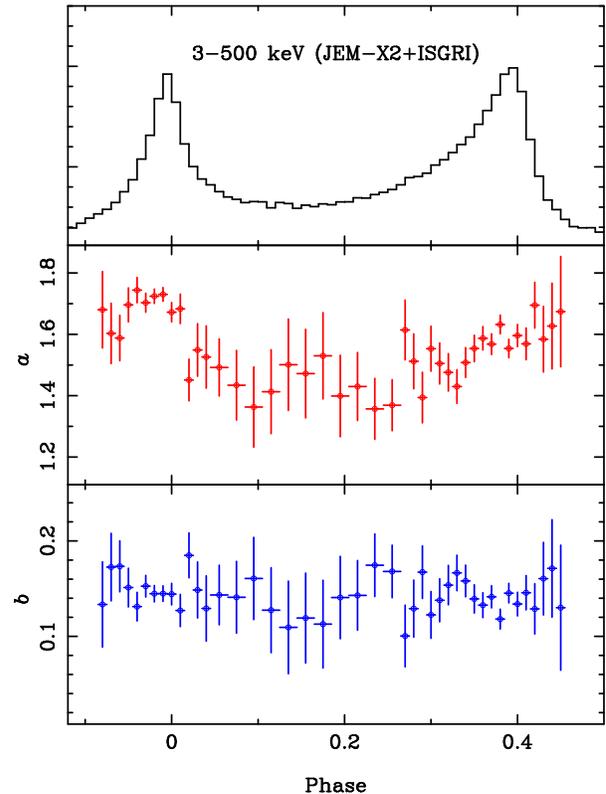} 
}}
\caption{Best fit parameters $a$ and $b$ measured fitting simultaneously 
JEM-X2 and IBIS/ISGRI with the curved model of Eq. (1) vs. phase}
\end{figure}

Phase histograms of the Crab pulsar were evaluated for each instrument and
each observation using the period folding technique and
adding the various  offsets quoted in \citet{walt03}
to correct the time relation derived by the INTEGRAL Science Data Center.
The resulting phase histograms in six energy bands from 3 keV to 360 keV are
shown in Fig. 1 in absolute phase with a phase resolution ranging from 
0.01 (0.33 ms) to 0.03 (1.1 ms) according to the available statistics and 
to the instrument time resolution. 
\\
The well-known double peaked structure is prominent in 
all the profiles with a high statistical significance
and the known evolution of the Crab pulse 
profile with  energy can be observed: the relative intensity of the first
pulse  respect to the second increases with energies together with the  
level of the bridge emission.


\section{Spectral Analysis}
Crab pulsar phase histograms  with  
 100 phase bins  for JEM-X1, JEM-X2 and IBIS/ISGRI  and 50 phase  bins for SPI
 were generated for each energy channel of each instrument and organized in an
energy-phase matrix.
A detailed phase resolved spectral analysis in the range ($-$0.1,$+$0.46)
was performed by 
selecting spectra in phase intervals 0.01 wide in the two main peaks
and 0.02 in the interpeak for JEM-X1, JEM-X2 and IBIS/ISGRI;
phase intervals of double size (0.02 at the two  peaks and 0.04 in the
interpeak) were considered for SPI data.
The nebular emission  and  the instrumental background for each
phase resolved spectra were subtracted evaluating them from 
 the off-pulse level in the pase 
 interval ($+$0.60,$+$0.80) and considering that the
 contribution of the pulsed emission in this phase interval is negligible
\citep{weis04}.  

Energy channels are uniformly rebinned in
agreement with the response matrices and in order 
to have a minimum bin content of 20 counts.

Spectra of each instrument were first modeled  with a single power 
law; low energy absorption has been included in the JEM-X fits, fixing the
absorbing column to the values derived by \citet{weis04}. 
Spectra from the two JEM-X units were fitted simultaneously introducing
as free parameter a  factor to take care of the instrumental systematics.
The best fit values of this intercalibration factor are in the range
0.8-1.06 with an average of 0.98.

JEM-X reduced $\chi^2$ are generally acceptable: they span the range 0.8 1.2
(d.o.f 266) with only four values  out of 42 above 
this range (between 1.29 and 1.33). 

IBIS/ISGRI fits gave values of reduced $\chi^2$ between 0.8 and 1.8 (122 dof)
with 16 values greater than 1.26 over 42 and only one above 1.5. 
These higher values of reduced $\chi^2$ can be considered acceptable
because they  are  due to local
residuals at low  energies as expected from the level of accuracy
of the systematics in OSA 4.2 software\footnote{see the report
osa\_sci\_val\_isgri-1.0.pdf available at \\
\texttt{http://isdc.unige.ch/Soft/download/osa/osa\_doc/prod}}.
The SPI values of the reduced  $\chi^2$  lie in the expected range
0.7-1.4 (41 d.o.f.) with only two values at 1.6. The best fit  
spectral indices  are shown  in Fig. 2 vs. phase together with the light 
curves for the three instruments.
As example in Fig. 3 (top panels) 
the JEM-X, IBIS/ISGRI and SPI spectra relative to the
 the first peak are shown together with the
residuals respect to the simple power-law model (bottom panels).

The same phase dependence is clearly apparent in each plot in Fig. 2: 
the first peak has the softest spectrum, whereas the hardest emission 
is produced in the interpeak.
The statistical significance of the softening 
of the spectral index has been evaluated fitting the photon index 
in the leading edge of the first peak with a constant and with a line. 
Applying the  F-test to the derived  $\chi^2$, a significance of 
99.6\% in the JEM-X2 energy range and 99.7\% in the IBIS/ISGRI 20-500 keV 
can be inferred confirming the results 
obtained by  \citet{prav97} and \citet{mass00}.

The presence of a line at 440 keV
in the ISGRI spectrum relative to the  phase interval (0.27--0.47)
has also been investigated.
We find a 3 $\sigma$ upper limit of  1.4$\times$10$^{-3}$ ph cm$^{-2}$
s$^{-1}$ consistent with the presence of the line detected by \citet{mass91}.

Comparing JEM-X, SPI and IBIS/ISGRI results, we
note that spectral indices are clearly increasing with
energy over all the phase intervals, in agreement with BeppoSAX results
\citep{mass00}.
It is already known that the spectral energy distribution of the Crab pulsed
emission is continously steepening from the optical frequencies to $\gamma$-rays. 
In the X-ray range, \citet{mass00} showed that a suitable model
is the curved power law with a continously steepening described by the 
following formula:
$$F(E)=K\,E^{-(a+b\,Log(E))} 
\eqno(1)$$
where {\it a} corresponds to the photon index at 1 keV and {\it b} measures 
the curvature of the spectral distribution. 

We fitted at first JEM-X2 and ISGRI spectra  simultaneously  with a single
power law introducing a normalization factor in the model
to take into account the intercalibration systematics between the two
instruments.
In this wide band fits we considered only JEM-X2 units that shows a better
calibration compared to the BeppoSAX-MECS. We find, in fact that the
discrepancies in the photon indices measured by the two instruments,
in the common energy range 2-10 keV, in the same phase intervals  
is lower than 5\%.  SPI has not been included  because of
the lower statistics.
The resulted reduced $\chi^2$ have values generally unacceptable
for the expected distribution with 256 degree of freedom. The distribution of
the  reduced $\chi^2$ is shown in the top panel of Fig. 4: values
$>$ 1.3 are relative to the spectra in the two main peaks.
Following \citet{mass00} approach, we fitted then  JEM-X2 and 
ISGRI spectra  simultaneously with the curved  model of Eq. (1).
All fits gave acceptable $\chi^2$, as shown in the bottom panel of Fig. 4 and
the best fit values of intercalibration factors are compatible with the
constant 0.92$\pm$0.01 over all phase intervals.
The best fit  values of the two parameters vs phase are shown in Fig. 5. 
The bending parameter {\it b} is statistically compatible with a single value
over all phase intervals as found by Massaro et al. (2001) over
3 wider intervals. The fit  with a constant gave a  value of 
0.14$\pm$0.02, where the
error  represents the spread around the average.
\\


\section{Discussion}
INTEGRAL observations of the Crab pulsar provided a high-statistical
data set for the  study of the spectral and phase distribution 
of the pulsed emission
over a wide  X-ray energy interval (3-500 keV). 
The main result from the analysis of 
these data is that the pulsed spectral distribution 
can be accurately represented by a curved function.
The values of the bending
parameter {\it b} in different phase intervals are 
consistent within the errors with a constant. 
This is a first independent confirmation of 
BeppoSAX results presented  in \citet{mass00}.
Other results of the timing and spectral analysis can be
summarized in the following points:
\begin{itemize}
\item 
the  spectral distribution 
changes with phase: the photon index softens towards P1, hardens in the Ip
region and increases again in the second peak with  a
characteristic reverse S shape over the 3-500 keV energy range;
\item
the photon index at the P1 leading edge shows a significative increase both in the
JEM-X2 and in the  IBIS/ISGRI  energy ranges
 confirming the results obtained by \citet{prav97} and \citet{mass00}
yet questioned by \citet{vive02};
\item
the analysis of IBIS/ISGRI spectrum relative to the phase interval
(0.27--0.47) does not rule out the presence of the 440 keV line detected by 
\citet{mass91}.
\end{itemize}

\noindent
The value of the bending parameter (0.14$\pm$0.02)  is
similar to the values obtained for other Crab-like pulsars
\citep{depl03,cusu01,mine04}
 strongly suggesting a common characteristic of these sources.
\\
A log-parabolic spectrum, can be interpreted in term of the physics of the particle
acceleration. 
It can be obtained when the acceleration  decreases with  the particle
energy \citet{mass04a,mass04b}. In the case of the pulsar environment, this could 
result from several crossings  of the magnetosphere gaps with a time of 
permanence inside the acceleration region that decrease with the energy of 
the particles. 
\\
The Spectral Energy Distribution (SED) of the log-parabolic law has a
maximum at the energy $E_p$ given by 
$$E_{p}=10^{(2-a)/2 \, b}
\eqno(2)
$$
Considering the variation with phase of the parameter $a$, the values of
$E_{p}$ ranges between $\sim$10 keV and $\sim$130 keV, 
in first peak and in the interpeak phase intervals, respectively.
A possible explanation of the  phase variation of the maximum energy is
that we are observing photons emitted at different levels of the
magnetosphere. 
The  three-dimensional outer gap model  \citep{chen00}
seems to provide a viable theoretical description. 
In the framework of this model, curvature gamma-rays are converted into 
electron-positron
pairs by the interaction with the magnetic field and X-ray photons are then
radiated by these secondary particles as synchrotron emission with a
typical photon energy $E_{syn}$:
$$E_{syn}(r)=\frac{3}{2}\,\left ( \frac{E_e}{mc^2} \right )^2 \,
\frac{h e\, B(r) \,{\rm sin} \beta(r)}{mc}
\eqno(3)
$$
\noindent
where $E_e$ is the electron energy $B(r)$ is the dipole magnetic field,  sin
$\beta(r)\propto (r)^{1/2}$
the pitch angle and $r$ the hight from the star surface within the magnetosphere.
Assuming that $E_e$   is  a fraction of the
curvature energy  estimated as:

$$E_{cur}(r)=\frac{3}{2} \hbar \gamma^3_e(r) \,\frac{c}{s(r)},
\eqno(4)
$$

\noindent
where $\gamma_e(r)\propto (r)^{-1/8}$ is the local Lorentz factor 
and $s(r)\propto (r)^{1/2}$ 
is the curvature radius,
the ratio between the two SED maxima can be related to the ratio of the
{\bf height} of the emission
regions $r_1$ and $r_2$:

$$\frac {E_{syn}(r_1)}{E_{syn}(r_2)}=\frac{E_{cur}(r_1)}{E_{cur}(r_2)} \,
\frac{B(r_1)}{B(r_2)} \, \frac{{\rm sin} \beta(r_1)}{{\rm sin} \beta(r_2)}.
\eqno(5)
$$

\noindent
Introducing the dependence from $r$ of each variable  we find 
the following simplified relation:  

$$\frac {E_{syn}(r_1)}{E_{syn}(r_2)} \simeq \left ( \frac{r_2}{r_1} \right )^{2.7}.
\eqno(6)
$$
\noindent
Results  from INTEGRAL spectral phase resolved analysis 
on the two average energies imply a
ratio of  the  heights of emitting regions of $\sim$ 0.3-0.4 in agreement with the
plot in Fig.9 of \citet{chen00} where the level of the emission region
vs phase computed for the Crab pulsar is shown.

The present analysis of the Crab X-ray pulsed emission confirms that 
a wide energy band analysis is very important for the study and understanding
of the SED of radio pulsars. Future mission sensitive to the $\gamma$ rays 
should include this source as a primary scientific target.

\begin{acknowledgements}
TM is grateful to Enrico Massaro for his helpful suggestions and 
discussion on the paper. The authors thanks the anonimous referee for his/her
relevant comments that greatly improved the scientific content of the paper.
\end{acknowledgements}

\appendix

\section{ISGRI energy correction and response matrix}

\begin{figure*}[htb]
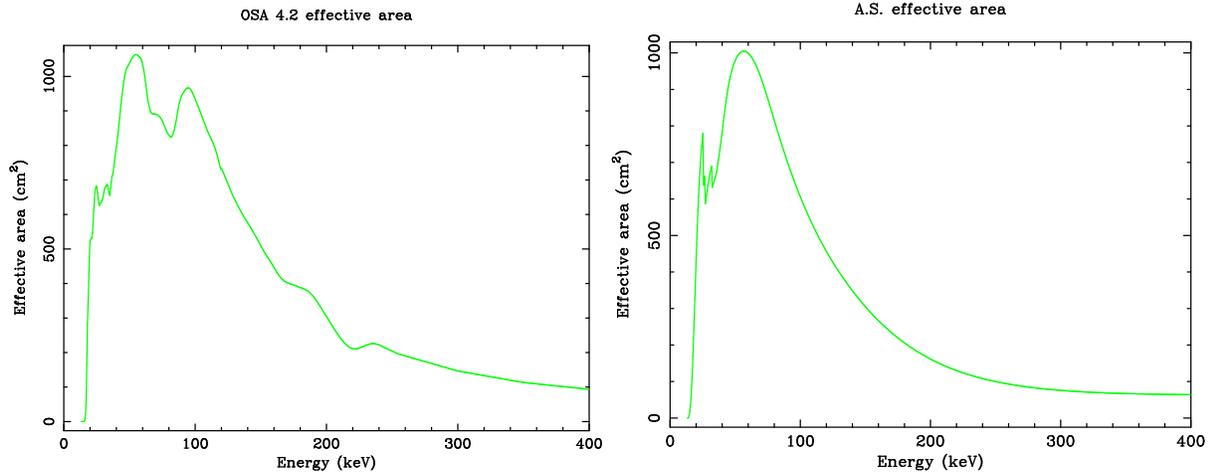

\label{fig6}
\centerline{ 
\hbox{
\psfig{figure=4305fa1a.ps,width=7.9cm,angle=-90,clip=} 
\psfig{figure=4305fa1b.ps,width=7.9cm,angle=-90,clip=} 
}}
\caption{ISGRI effective area vs energy for the standard software (left panel)
and for the  A.S. version (right panel). Features in the range 30-40 keV are 
due to CdTe absorption edges. }
\end{figure*}

The ISGRI detection layer, that consists of an 128$\times$128 array of
independent CdTe detector pixels, suffer of a rather severe
"Charge Loss Effect", common to this kind of detectors.
To take into account these effects,  the solution adopted by the calibration team
\citep{lebr03} is to perform an energy correction as a function of the pulse 
rise-time  using multiplicative coefficients stored in a "Look-Up" Table 
(LUT2). The version of the LUT2 distributed with the OSA software 
is not yet optimized\footnote{See A. Segreto 
talk at the Internal INTEGRAL workshop available at\\
\texttt{http://www.rssd.esa.int/Integral/workshops/Jan2005/}}, and it introduces 
artificial features in the spectra, the most relevant in the 80 keV region,
that are compensated with  
an ad-hoc modifications of the ISGRI  effective area\footnote{see
the report osa\_sci\_val\_isgri-1.0.pdf available at \\
\texttt{http://isdc.unige.ch/Soft/download/osa/osa\_doc/prod}}
(see left panel of Fig.~\ref{fig6}1).
However, the intensity of the artificial features strongly depend on  the
spectral shape and the analysis of 
sources with spectral shapes different from that of the Crab
might be affected.

\begin{figure*}[htb]
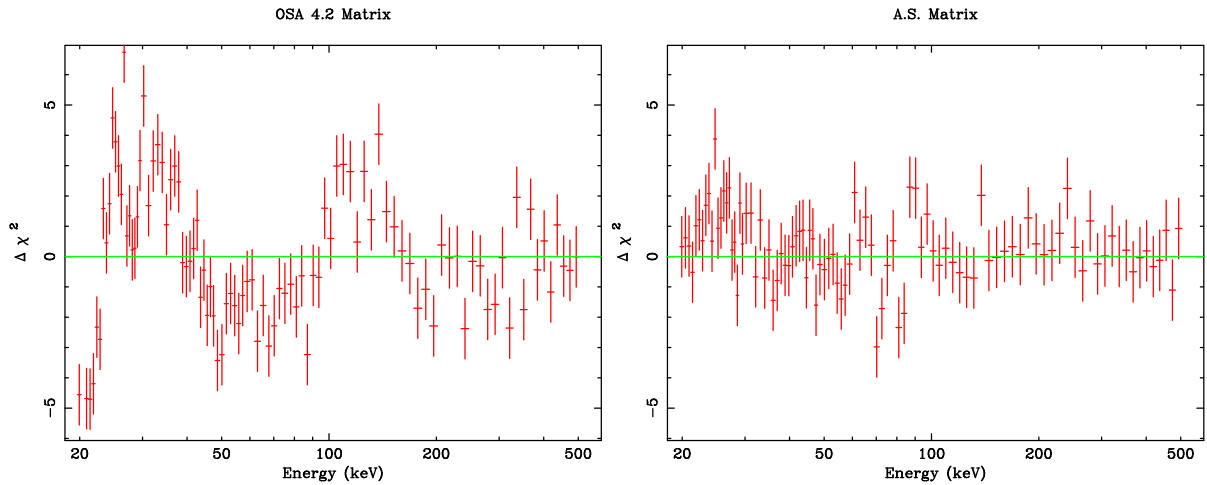

\label{fig7}
\centerline{ 
\hbox{
\psfig{figure=4305fa2a.ps,width=7.9cm,angle=-90,clip=}
\psfig{figure=4305fa2b.ps,width=7.9cm,angle=-90,clip=} 
}}
\caption{ {\it Left Panel}: Residuals of the power law fit of the Crab spectrum 
obtained processing ISGRI data with the standard LUT2 and
 effective area. {\it Right Panel}: Residuals of the power law fit of the Crab 
spectrum obtained processing ISGRI data with the  A.S. LUT2 and  effective area  }
\end{figure*}

A new LUT2\footnote{the file is  available from the web page\\
\texttt{http://www.ifc.inaf.it/\~{}ferrigno/integral/ISGRI\_alternative\_IC}} 
based on on-ground and in-flight calibration data has been
generated by one of the authors A.Segreto.
The better energy correction is confirmed by the fact that it is no more necessary to
introduce ad-hoc wiggles in the effective area, as shown in the right panel of
Fig.~\ref{fig6}1.  Moreover, the new effective area gives a value of
the Crab spectral index in better agreement with the one quoted in literature 
and measured by the other instrument on-board INTEGRAL\footnote{see the report
  osa\_cross\_cal-1.0.pdf available at \\
\texttt{ http://isdc.unige.ch/Soft/download/osa/osa\_doc/prod/}
} (see also Table~\ref{tab2}1).
In Fig.~\ref{fig7}2, the residuals of the power law fit of the Crab spectrum 
obtained processing the ISGRI data with the standard OSA 4.2 response matrix
(left panel) and with the A.S. LUT2 and effective area (right panel) are shown.

Tests on this matrix have been performed analysing sources which are 
detected with good statistics up to 100--200\,keV.
The spectral parameter derived with the two matrices are plotted in 
Table~\ref{tab2}1 together with the values quoted in literature.
The matrix we adopted gives best fit values of the spectral parameters in
agreement with  the ones quoted in literature and
$\chi^2$ values  generally lower than the one derived from the standard
OSA 4.2 matrix.

\begin{table*}[h]
\label{tab2} 
\caption{Comparison of the best fit parameters obtained using the standard
  matrix (ISDC) and the new A.S. version.}
\begin{center}
\begin{tabular}{@{}llll}
\hline 
\hline
Parameter & OSA 4.2 & A.S. Version & Other experiment \\ 
\hline
\multicolumn{3}{l}{{\it {\bf Crab nebula+pulsar}: Power Law }} \\
$\alpha$           & 2.19$\pm$0.03$^*$     & 2.09$\pm$0.03  & 2.1  \\
$K$  (ph cm$^{-2}$ s$^{-1}$ KeV$^{-1}$)    &  10.2$\pm$0.2$^*$     & 9.6$\pm$0.2 & 9.7 \\
$\chi^2$ (d.o.f.)  &  5.6  (95)            & 1.3 (95) & \\
                   &       &          & Ref (1) \\
\hline
\multicolumn{4}{l}{{\it {\bf \object{GX 1+4}}: Comptonization continuum model (COMPTT in XSPEC)}} \\
$T_0$ (keV)  &   1.27 (frozen)   &  1.27 (frozen) &   1.27$^{+1.4}_{-1.1}$   \\
$T_e$(keV)   & 13.5$\pm$0.5      &  12.6$\pm$0.4  &   10.3$^{+13}_{-9}$  \\
$\tau$     &  2.7$\pm$0.1      &  3.3$\pm$0.2   &   3.7$^{+4}_{-3}$ \\
$K$ (ph cm$^{-2}$ s$^{-1}$ KeV$^{-1}$)& (2.1$\pm$0.1)$\times$10$^{-2}$ & (1.9$\pm$0.1)$\times$10$^{-2}$ & 2.1$^{+2.3}_{-1.8}\times$10$^{-2}$  \\
$\chi^2$ (d.o.f.) & 1.5 (58) & 1.1 (58)  & \\
                   &       &          & Ref (2) \\
\hline
\multicolumn{4}{l}{{\it  {\bf\object{ Cygnus X1}}: Cut-off power law}} \\
$\alpha$      & 1.66$\pm$0.04$^*$     &  1.45$\pm$0.04    &  1.45$\pm$0.01  \\
E$_c$ (keV)   & 158$\pm$12$^*$        &  122$\pm$7        &  162$\pm$9 \\
$K$ (ph cm$^{-2}$ s$^{-1}$ KeV$^{-1}$)    & 1.7$\pm$0.3$^*$  &  1.2$\pm$0.3   &  0.91$\pm$0.04 \\
$\chi^2$ (d.o.f.) & 3.2 (48) & 1.9  (48) \\
                   &       &          & Ref (3) \\
\hline
\multicolumn{4}{l}{{\it  {\bf \object{Sco X1}}: Bremstralung + Power law}} \\
$KT$ (keV)                                    & 4.2$\pm$0.1                   & 4.3$\pm$0.1  & 4.51$\pm$0.08   \\
$Flux_{20-50 keV}$ (erg cm$^{-2}$ s$^{-1}$)   &(5.5$\pm$0.3)$\times$10$^{-9}$ & (6.5$\pm$0.4)$\times$10$^{-9}$ &(7.4$\pm$0.6)$\times$10$^{-9}$ \\
$\alpha$                                      & 2.9$\pm$0.2                  & 2.8$\pm$0.2   &  2.4$\pm$0.3 \\
$Flux_{20-200 keV}$ [erg cm$^{-2}$ s$^{-1}$]  &(0.8$\pm$0.4)$\times$10$^{-9}$ & (0.7$\pm$0.4)$\times$10$^{-9}$ &(1.04$\pm$0.08)$\times$10$^{-9}$ \\
$\chi^2$ (d.o.f.)                             & 1.3 (52)                      & 1.4 (52) &\\
                   &       &          & Ref (4) \\
\hline
\end{tabular}
\end{center}
$^*$ Errors are computed introducing systematic errors in
  order to reduce the $\chi^2$ below 2.0 \\
(1) \citet{torr74};  
(2) \citet{gall00}; 
(3) \citet{dove98}; 
(4) \citet{dami01}
\end{table*}

\bibliographystyle{aa}
\bibliography{bibliografia_pulsar}

\begin{thebibliography}{37}
\expandafter\ifx\csname natexlab\endcsname\relax\def\natexlab#1{#1}\fi

\bibitem[{{Brandt} {et~al.}(2003){Brandt}, {Budtz-J{\o}rgensen}, {Lund},
  {Rasmussen}, {Laursen}, {Chenevez}, {Westergaard}, {Juchnikowski}, {Walter},
  {Schmidt}, \& {Much}}]{bran03}
{Brandt}, S., {Budtz-J{\o}rgensen}, C., {Lund}, N., {et~al.} 2003, \aap, 411,
  L433

\bibitem[{{Cheng} {et~al.}(2000){Cheng}, {Ruderman}, \& {Zhang}}]{chen00}
{Cheng}, K.~S., {Ruderman}, M., \& {Zhang}, L. 2000, \apj, 537, 964

\bibitem[{{Cusumano} {et~al.}(2001){Cusumano}, {Mineo}, {Massaro}, {Nicastro},
  {Trussoni}, {Massaglia}, {Hermsen}, \& {Kuiper}}]{cusu01}
{Cusumano}, G., {Mineo}, T., {Massaro}, E., {et~al.} 2001, \aap, 375, 397

\bibitem[{{D'Amico} {et~al.}(2001){D'Amico}, {Heindl}, {Rothschild}, \&
  {Gruber}}]{dami01}
{D'Amico}, F., {Heindl}, W.~A., {Rothschild}, R.~E., \& {Gruber}, D.~E. 2001,
  \apjl, 547, L147

\bibitem[{{de Plaa} {et~al.}(2003){de Plaa}, {Kuiper}, \& {Hermsen}}]{depl03}
{de Plaa}, J., {Kuiper}, L., \& {Hermsen}, W. 2003, \aap, 400, 1013

\bibitem[{{Di Cocco} {et~al.}(2003){Di Cocco}, {Caroli}, {Celesti}, {Foschini},
  {Gianotti}, {Labanti}, {Malaguti}, {Mauri}, {Rossi}, {Schiavone},
  {Spizzichino}, {Stephen}, {Traci}, \& {Trifoglio}}]{dico03}
{Di Cocco}, G., {Caroli}, E., {Celesti}, E., {et~al.} 2003, \aap, 411, L189

\bibitem[{{Dove} {et~al.}(1998){Dove}, {Wilms}, {Nowak}, {Vaughan}, \&
  {Begelman}}]{dove98}
{Dove}, J.~B., {Wilms}, J., {Nowak}, M.~A., {Vaughan}, B.~A., \& {Begelman},
  M.~C. 1998, \mnras, 298, 729

\bibitem[{{Galloway}(2000)}]{gall00}
{Galloway}, D.~K. 2000, \apjl, 543, L137

\bibitem[{{Goldwurm} {et~al.}(2003){Goldwurm}, {David}, {Foschini}, {Gros},
  {Laurent}, {Sauvageon}, {Bird}, {Lerusse}, \& {Produit}}]{gold03}
{Goldwurm}, A., {David}, P., {Foschini}, L., {et~al.} 2003, \aap, 411, L223

\bibitem[{{Gros} {et~al.}(2003){Gros}, {Goldwurm}, {Cadolle-Bel}, {Goldoni},
  {Rodriguez}, {Foschini}, {Del Santo}, \& {Blay}}]{gros03}
{Gros}, A., {Goldwurm}, A., {Cadolle-Bel}, M., {et~al.} 2003, \aap, 411, L179

\bibitem[{{Kuiper} {et~al.}(2001){Kuiper}, {Hermsen}, {Cusumano}, {Diehl},
  {Sch{\" o}nfelder}, {Strong}, {Bennett}, \& {McConnell}}]{kuip01}
{Kuiper}, L., {Hermsen}, W., {Cusumano}, G., {et~al.} 2001, \aap, 378, 918

\bibitem[{{Kuiper} {et~al.}(2003){Kuiper}, {Hermsen}, {Walter}, \&
  {Foschini}}]{kuip03}
{Kuiper}, L., {Hermsen}, W., {Walter}, R., \& {Foschini}, L. 2003, \aap, 411,
  L31

\bibitem[{{Lebrun} {et~al.}(2003){Lebrun}, {Leray}, {Lavocat}, {Cr{\' e}tolle},
  {Arqu{\` e}s}, {Blondel}, {Bonnin}, {Bou{\` e}re}, {Cara}, {Chaleil}, {Daly},
  {Desages}, {Dzitko}, {Horeau}, {Laurent}, {Limousin}, {Mathy}, {Mauguen},
  {Meignier}, {Molini{\' e}}, {Poindron}, {Rouger}, {Sauvageon}, \&
  {Tourrette}}]{lebr03}
{Lebrun}, F., {Leray}, J.~P., {Lavocat}, P., {et~al.} 2003, \aap, 411, L141

\bibitem[{{Lund} {et~al.}(2003){Lund}, {Budtz-J{\o}rgensen}, {Westergaard},
  {Brandt}, {Rasmussen}, {Hornstrup}, {Oxborrow}, {Chenevez}, {Jensen},
  {Laursen}, {Andersen}, {Mogensen}, {Rasmussen}, {Om{\o}}, {Pedersen},
  {Polny}, {Andersson}, {Andersson}, {K{\" a}m{\" a}r{\" a}inen}, {Vilhu},
  {Huovelin}, {Maisala}, {Morawski}, {Juchnikowski}, {Costa}, {Feroci},
  {Rubini}, {Rapisarda}, {Morelli}, {Carassiti}, {Frontera}, {Pelliciari},
  {Loffredo}, {Mart{\'{\i}}nez N{\' u}{\~ n}ez}, {Reglero}, {Velasco},
  {Larsson}, {Svensson}, {Zdziarski}, {Castro-Tirado}, {Attina}, {Goria},
  {Giulianelli}, {Cordero}, {Rezazad}, {Schmidt}, {Carli}, {Gomez}, {Jensen},
  {Sarri}, {Tiemon}, {Orr}, {Much}, {Kretschmar}, \& {Schnopper}}]{lund03}
{Lund}, N., {Budtz-J{\o}rgensen}, C., {Westergaard}, N.~J., {et~al.} 2003,
  \aap, 411, L231

\bibitem[{{Massaro} \& {Cusumano}(2003)}]{mass03}
{Massaro}, E. \& {Cusumano}, G. 2003, in Pulsars, AXPs and SGRs Observed with
  BeppoSAX and Other Observatories, 15--22

\bibitem[{{Massaro} {et~al.}(2000){Massaro}, {Cusumano}, {Litterio}, \&
  {Mineo}}]{mass00}
{Massaro}, E., {Cusumano}, G., {Litterio}, M., \& {Mineo}, T. 2000, \aap, 361,
  695

\bibitem[{{Massaro} {et~al.}(2001){Massaro}, {Litterio}, {Cusumano}, \&
  {Mineo}}]{mass01}
{Massaro}, E., {Litterio}, M., {Cusumano}, G., \& {Mineo}, T. 2001, in ESA
  SP-459: Exploring the Gamma-Ray Universe, 229--233

\bibitem[{{Massaro} {et~al.}(1991){Massaro}, {Matt}, {Salvati}, {Costa},
  {Mandrou}, {Niel}, {Olive}, {Mineo}, {Sacco}, {Scarsi}, {Gerardi},
  {Agrinier}, {Barouch}, {Comte}, {Parlier}, \& {Masnou}}]{mass91}
{Massaro}, E., {Matt}, G., {Salvati}, M., {et~al.} 1991, \apjl, 376, L11

\bibitem[{{Massaro} {et~al.}(2004{\natexlab{a}}){Massaro}, {Perri}, {Giommi},
  \& {Nesci}}]{mass04a}
{Massaro}, E., {Perri}, M., {Giommi}, P., \& {Nesci}, R. 2004{\natexlab{a}},
  \aap, 413, 489

\bibitem[{{Massaro} {et~al.}(2004{\natexlab{b}}){Massaro}, {Perri}, {Giommi},
  {Nesci}, \& {Verrecchia}}]{mass04b}
{Massaro}, E., {Perri}, M., {Giommi}, P., {Nesci}, R., \& {Verrecchia}, F.
  2004{\natexlab{b}}, \aap, 422, 103

\bibitem[{{Mineo} {et~al.}(2004){Mineo}, {Cusumano}, \& {Massaro}}]{mine04}
{Mineo}, T., {Cusumano}, G., \& {Massaro}, E. 2004, Nuclear Physics B
  Proceedings Supplements, 132, 632

\bibitem[{{Mineo} {et~al.}(1997){Mineo}, {Cusumano}, {Segreto}, {Massaro}, {dal
  Fiume}, {Giarrusso}, {Matteuzzi}, {Nicastro}, \& {Parmar}}]{mine97}
{Mineo}, T., {Cusumano}, G., {Segreto}, A., {et~al.} 1997, \aap, 327, L21

\bibitem[{{Pravdo} {et~al.}(1997){Pravdo}, {Angelini}, \& {Harding}}]{prav97}
{Pravdo}, S.~H., {Angelini}, L., \& {Harding}, A.~K. 1997, \apj, 491, 808

\bibitem[{{Rots} {et~al.}(2004){Rots}, {Jahoda}, \& {Lyne}}]{rots04}
{Rots}, A.~H., {Jahoda}, K., \& {Lyne}, A.~G. 2004, \apjl, 605, L129

\bibitem[{{Skinner} \& {Connell}(2003)}]{skin03}
{Skinner}, G. \& {Connell}, P. 2003, \aap, 411, L123

\bibitem[{{Strong}(2003)}]{stro03}
{Strong}, A.~W. 2003, \aap, 411, L127

\bibitem[{{Tennant} {et~al.}(2001){Tennant}, {Becker}, {Juda}, {Elsner},
  {Kolodziejczak}, {Murray}, {O'Dell}, {Paerels}, {Swartz}, {Shibazaki}, \&
  {Weisskopf}}]{tenn01}
{Tennant}, A.~F., {Becker}, W., {Juda}, M., {et~al.} 2001, \apjl, 554, L173

\bibitem[{{Toor} \& {Seward}(1974)}]{torr74}
{Toor}, A. \& {Seward}, F.~D. 1974, \aj, 79, 995

\bibitem[{{Ubertini} {et~al.}(2003){Ubertini}, {Lebrun}, {Di Cocco}, {Bazzano},
  {Bird}, {Broenstad}, {Goldwurm}, {La Rosa}, {Labanti}, {Laurent}, {Mirabel},
  {Quadrini}, {Ramsey}, {Reglero}, {Sabau}, {Sacco}, {Staubert}, {Vigroux},
  {Weisskopf}, \& {Zdziarski}}]{uber03}
{Ubertini}, P., {Lebrun}, F., {Di Cocco}, G., {et~al.} 2003, \aap, 411, L131

\bibitem[{{Ulmer} {et~al.}(1994){Ulmer}, {Lomatch}, {Matz}, {Grabelsky},
  {Purcell}, {Grove}, {Johnson}, {Kinzer}, {Kurfess}, {Strickman}, {Cameron},
  \& {Jung}}]{ulme94}
{Ulmer}, M.~P., {Lomatch}, S., {Matz}, S.~M., {et~al.} 1994, \apj, 432, 228

\bibitem[{{Vedrenne} {et~al.}(2003){Vedrenne}, {Roques}, {Sch{\" o}nfelder},
  {Mandrou}, {Lichti}, {von Kienlin}, {Cordier}, {Schanne}, {Kn{\" o}dlseder},
  {Skinner}, {Jean}, {Sanchez}, {Caraveo}, {Teegarden}, {von Ballmoos},
  {Bouchet}, {Paul}, {Matteson}, {Boggs}, {Wunderer}, {Leleux},
  {Weidenspointner}, {Durouchoux}, {Diehl}, {Strong}, {Cass{\' e}}, {Clair}, \&
  {Andr{\' e}}}]{vedr03}
{Vedrenne}, G., {Roques}, J.-P., {Sch{\" o}nfelder}, V., {et~al.} 2003, \aap,
  411, L63

\bibitem[{{Vivekanand}(2002)}]{vive02}
{Vivekanand}, M. 2002, \aap, 391, 1033

\bibitem[{{Walter} {et~al.}(2003){Walter}, {Favre}, {Dubath}, {Domingo},
  {Gienger}, {Hermsen}, {Pineiro}, {Kuiper}, {Schmidt}, {Skinner}, {Tuttlebee},
  {Ziegler}, \& {Courvoisier}}]{walt03}
{Walter}, R., {Favre}, P., {Dubath}, P., {et~al.} 2003, \aap, 411, L25

\bibitem[{{Weisskopf} {et~al.}(2004){Weisskopf}, {O'Dell}, {Paerels}, {Elsner},
  {Becker}, {Tennant}, \& {Swartz}}]{weis04}
{Weisskopf}, M.~C., {O'Dell}, S.~L., {Paerels}, F., {et~al.} 2004, \apj, 601,
  1050

\bibitem[{{Westergaard} {et~al.}(2003){Westergaard}, {Kretschmar}, {Oxborrow},
  {Larsson}, {Huovelin}, {Maisala}, {Mart{\'{\i}}nez N{\'u}{\~n}ez}, {Lund},
  {Hornstrup}, {Brandt}, {Budtz-J{\o}rgensen}, \& {Rasmussen}}]{west03}
{Westergaard}, N.~J., {Kretschmar}, P., {Oxborrow}, C.~A., {et~al.} 2003, \aap,
  411, L257

\bibitem[{{Winkler} {et~al.}(2003){Winkler}, {Courvoisier}, {Di Cocco},
  {Gehrels}, {Gim{\' e}nez}, {Grebenev}, {Hermsen}, {Mas-Hesse}, {Lebrun},
  {Lund}, {Palumbo}, {Paul}, {Roques}, {Schnopper}, {Sch{\" o}nfelder},
  {Sunyaev}, {Teegarden}, {Ubertini}, {Vedrenne}, \& {Dean}}]{wink03}
{Winkler}, C., {Courvoisier}, T.~J.-L., {Di Cocco}, G., {et~al.} 2003, \aap,
  411, L1

\bibitem[{{Zhang} \& {Cheng}(2002)}]{zhan02}
{Zhang}, L. \& {Cheng}, K.~S. 2002, \apj, 569, 872

\end{thebibliography}

\end{document}